\documentclass{epl}
\title{Two-phonon pseudogap in the Klein-Gordon lattice}
\usepackage{rotating,texdraw}
\author{Laurent Proville}

\institute{Service de Recherches de M\'etallurgie Physique,
CEA-Saclay/DEN/DMN\\
91191-Gif-sur-Yvette Cedex, France }


\begin{document}

\maketitle

\date{\today}

\begin{abstract}
The energy spectrum of the quantum Klein-Gordon (KG) lattice is
computed numerically for model parameters relevant to optical
phonon spectra. A pairing of phonon states is found when
nonlinearity is significant, which agrees with other studies on
different quantum lattice models~\cite{AGRA,Eilbeck}. It results
from the lattice anharmonicity, the magnitude of which is
quantified by the binding energy of phonon bound states. Our work
focuses on the case of weak anharmonicity, i.e., the phonon
binding energy is weaker than the single phonon band width. We
find that the phonon pairs dissociate at the center of the lattice
Brillouin zone, whereas at the edge the binding energy remains
comparable to the width of the single phonon band. Consequently, a
weak nonlinearity is characterized by a pseudogap in the energy
spectrum of two-phonon states.


\end{abstract}

The phonon frequencies are some fundamental quantities that
provide information about atomic interactions and local structure
in crystals as well as in polymers or proteins. At the atomic
scale, the phonon frequencies are formally derived from the
quadratic expansion of the potential energy with respect to atomic
displacements. Actually a more accurate description of the
potential energy might be obtained by a higher order expansion
which involves some non-quadratic terms. Those terms are usually
referred to as nonlinear terms because they yield some forces that
are not proportional to displacements. In the early seventies, the
non-quadratic contribution to the energy has been quantified in
molecular crystals such as $CO_2$, $N_2 O$ and $OCS$ (see
Ref.~\cite{Bogani} and Refs. therein). The nonlinearity was then
identified by some anharmonic peaks in the infrared spectrum. In
solid $H_2$, similar infrared resonances were earlier
interpreted~\cite{Gush1957} as a signature of phonons pairing.
Currently the list of materials in which phonon bound states occur
is still growing~\cite{Jakob,Swanson}. In addition, some
controversial cases, as for instance acetanilide~\cite{Elder} or
the spectral branch $\Delta_2'$(LO) of
diamond~\cite{Ruvalds,Tubino} can be mentioned as still open
questions. In the present study, we point out a possible
difficulty for the vibration spectrum interpretation which could
occur when nonlinearity is not sufficiently strong to separate the
anharmonic resonances from the harmonic energy regions. Then, the
hybridization between bound and unbound phonons is shown to imply
a pseudogap in the lattice energy spectrum.

The anharmonicity of molecular crystals has been theoretically
studied by V.M. Agranovich~\cite{AGRA}, who proposed a qualitative
approach where some boson quasi-particles model the atom
vibrations. A Hubbard onsite interaction between boson pairs was
introduced to simulate the nonlinear behavior of the lattice (for
recent studies see Ref.~\cite{Eilbeck,Pouthier}). This effective
boson model displayed the concepts of phonon bound states and more
specifically of biphonon. The drawback of the model is that the
energy terms that do not conserve the boson number are neglected,
although those terms stem from the potential energy of atoms (or
molecules).
A rigorous approach to models which do not have a conserved boson
number was proposed in Ref.\cite{Mackay2000}. The present paper
can be considered as numerical support for that theory. Here, a
numerical method is introduced for computing the KG energy
spectrum with different forms of nonlinearity, i.e., different
orders in the potential energy expansion. Our technique is
based on the construction of a nonlinear phonon basis which allows
to treat lattices whatever the magnitude of nonlinearity is, in
contrast to the linear phonon basis~\cite{bishop96}, derived from
the harmonic approximation. The attractive advantage of our method
is that the maximum size of the lattices that can be studied is
large enough to approach the infinite system features, i.e., 
the energy spectrum shows no
boundary effects and it is nearly continuous in the reciprocal space.

The KG model may be introduced as follows. At the node $i$ of a
translational invariant d-dimensional lattice, the internal mode
$x_i$ of a unit cell (a molecule, for instance) evolves in the
local potential $V$ with an effective mass $m$, each of the $x_i$
being coupled to the nearest neighbors' modes by a quadratic
coupling, parametrized by $c$. Then, the Hamiltonian of that
system reads:
\begin{equation}
H = \sum_i [\frac{p_i^2}{2m} + V(x_i) - c \sum_{j=<i>}
(x_i-x_{j})^2] . \label{DebyeModif}
\end{equation}
The subscript $j=<i>$ denotes the first neighbors of site $i$. The
operator $x_i$ and its conjugate momentum $p_i$ satisfy $[x_i,p_i
]=i \hbar$. For small amplitudes of $x_i$, the Taylor expansion of
$V$ gives $V(x_i)=a_2 x_i^2 +a_3 x_i^3 +a_4 x_i^4$. It has been
limited to the fourth order but higher order terms can be taken
into account with no difficulties in calculations that follow.
Higher order coupling terms have been found not to modify
qualitatively our results. Introducing the dimensionless
coefficients and operators:
\begin{eqnarray}
A_3&= &a_3 \sqrt{\frac{\hbar}{m^3 \Omega^5}} \texttt{,} \ A_4= a_4
\frac{\hbar}{m^2 \Omega^3}, \
C= \frac{4 c}{m \Omega^2} \nonumber\\
P_i &=& p_i / \sqrt{m \hbar\Omega} \ \ \texttt{and} \  \ X_i=x_i
\sqrt{m\Omega/\hbar} \label{param}
\end{eqnarray}
where the frequency $\Omega = \sqrt{2 (a_2-(2d) c)/m}$
characterizes the $d$-dimensional oscillator networks, the
Hamiltonian reads:
\begin{equation}
H = \hbar \Omega \sum_i [\frac{P_i^2}{2} + \frac{X_i^2}{2} + A_3
X_i^3 + A_4 X_i^4 + \frac{C}{2}  X_i \sum_{j=<i>} X_{j}] .
\label{Hamilton}
\end{equation}
Note that the coupling coefficient~$c$ contributes to $\Omega$.
The width of an optical phonon band is physically a few percent of
the elementary excitation energy ($\approx \hbar \Omega$). We
restrict our study to $A_4\geq 0$ to ensure that the local
potential $V$ is positive when the onsite displacement increases
arbitrarily. In contrast, $A_3$ can take positive or negative
values provided $V$ is a monotonic function.

\begin{figure}
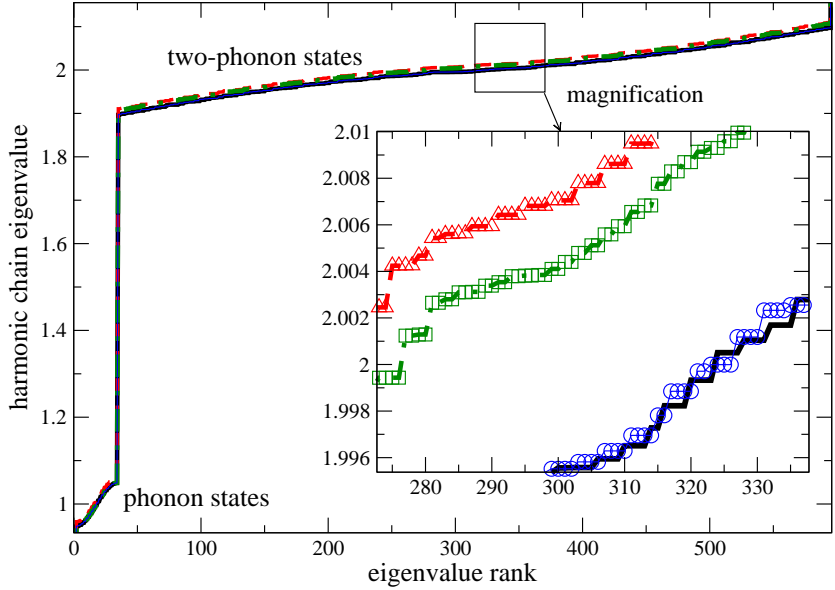

\onefigure[width=11cm]{Fig1EPL2004PROVILLE.eps} \caption{ For a
dimensionless coupling $C=0.05$, comparison between the exact
calculation (thick solid line) and our evaluations
(detailed in the text) of the eigenvalues in a 1D harmonic lattice
composed of
$S=33$ oscillators. Our numerics are performed for 3
distinct cutoffs $N_{cut} = 2$ (dashed line, triangles), $N_{cut} =3$
(dot-dashed line, squares) and $N_{cut} =4$ (thin solid line, circles).
The spectrum is measured with respect to the
groundstate and is arranged in increasing order. The eigenstate
number is reported on the X-axis.  The Y-axis
unit is $\hbar \Omega$.} \label{fig1}
\end{figure}

Our computation is now described. The first step is to
calculate eigenstates of the onsite Hamiltonian $h_i= P_i^2/2 +
X_i^2/2 +A_3 X_i^3 +A_4 X_i^4$, which is performed by projecting
$h_i$ over the usual Einstein basis. The diagonalization of the
corresponding matrix is realized numerically with great accuracy.
The eigenstates of $h_i$ are arranged by increasing order of
eigenvalues, so the $\alpha^{th}$ eigenstate is denoted
$\phi_{\alpha,i}$ and its eigenvalue is $\gamma(\alpha)$.
Considering the Hamiltonian $H_0= \hbar \Omega \sum_i h_i$, the
$H_0$ eigenstates can be written as products of onsite states
$\Pi_i\phi_{\alpha_i,i}$. Taking advantage of the lattice
translational degeneracy, some Bloch waves are constructed from
those products as follows:
\begin{eqnarray}
B_{[ \Pi_i \alpha_i]} (q) = \frac{1}{\sqrt{A_{ [ \Pi_i
\alpha_i]}}} \sum_j e^{-i q \times j a_0} \Pi_i \phi_{\alpha_i,i-j}
\label{OSPBW}
\end{eqnarray}
where $a_0$ is the lattice parameter, $q$ is the wave vector, the
subscript $[ \Pi_i \alpha_i]$ identifies different types of
products and $A_{[\Pi_i \alpha_i]}$ ensures the normalization.
Expanding the Hamiltonian $H$ in the Bloch waves  
is done analytically and gives a matrix $\cal B$$(q)$.
Since $H$ does not hybridize Bloch waves with different $q$, the
diagonalization of $\cal B$$(q)$ can be performed independently
for each $q$. To that aim, we used exact numerical methods (LAPACK
or Numerical Recipes) on a simple desktop PC. Then, the
Schr\"{o}dinger equation is solved with an error which shrinks to
zero exponentially by increasing the cutoff over the $H_0$
eigenstates. The cutoff is fixed by $\sum_i \alpha_i < N_{cut}$.
The number of Bloch waves involved in diagonalization is $3052$
for a lattice size $S=33$. The accuracy of our calculations has
been tested both on the anharmonic atomic chain $S=4$
\cite{bishop96} and the harmonic chain for which the eigenvalue
problem is solved analytically by a spatial Fourier transform of
Eq.~\ref{DebyeModif}. For these two comparisons, very good
agreements have been found. For the latter case, the results are
reported in Fig.~\ref{fig1}. Some distinct cutoff values $N_{cut}$
have been tested and clearly our eigenvalue evaluation converges
to the exact values as the cutoff increases. For two-phonon
excitations, the error is less than $0.1 \%$ when $N_{cut}=4$.
Then the accuracy is even better for the phonon band which
fits the dispersion law: $\hbar \Omega\sqrt{1+2C \cos(q a_0)}$.

\begin{figure}
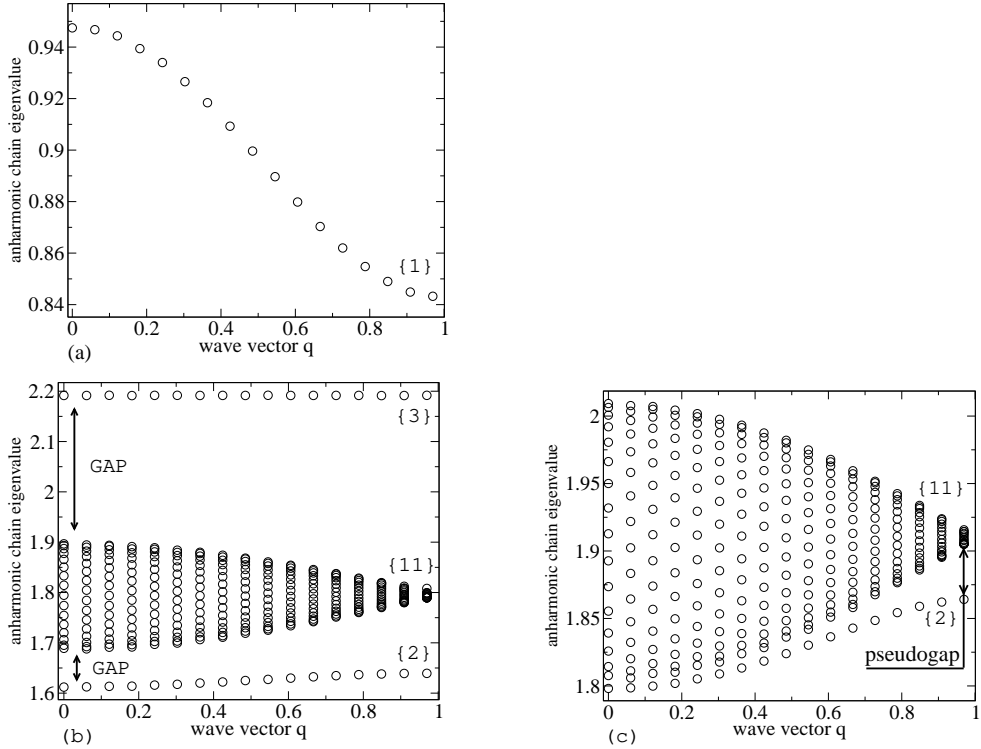

\oneimage[width=6.cm]{Fig2aEPL2004PROVILLE.eps}\vspace{0.2cm}
\twoimages[width=5.8cm]{Fig2bEPL2004PROVILLE.eps}{Fig2cEPL2004PROVILLE.eps}
\caption{Energy spectrum of a chain composed of $S=33$ unit cells:
(a) the optical phonon branch for model parameters $C=0.05$,
$A_3=0.13$ and $A_4=0.01$; (b) the two-phonon energy region for
same parameters as (a); (c) same as
(b) but $A_3=0.1$. The biphonon branch is labelled by $\{2\}$, the
unbound two-phonon bands by $\{11\}$ and the triphonon branch by
$\{3\}$. The Y-axis unit is $\hbar \Omega$ and its zero is
groundstate energy. The wave vector $q$ is reported on the X-axis,
whose unit is $(\pi/a_0)$. It ranges from the center to the edge
of the first lattice Brillouin zone.} \label{fig5}
\end{figure}

Our results are first presented for a 1D lattice. In
Figs.~\ref{fig5}(a-c), for a finite size $S=33$, the
eigen-energies (circle symbols) are plotted versus the wave
momentum $q$. They contribute to the distinct branches of the
spectrum. When the non-quadratic part of the energy is negligible,
the eigen-spectrum is composed of the fundamental optical branch
and the linear superpositions of phonon states. Increasing
gradually $A_4$, fixing $C=0.05$ (relevant for optical phonons,
see Fig.~\ref{fig5}(a)) and $A_3=0$, makes a branch split from the
top of the two-phonon band. Then, the onsite potential $V$ is said
to harden. Subsequently, if $A_4$ is fixed to a positive arbitrary
value and $A_3$ is increased in absolute value, the isolated
branch backs into the two-phonon band and eventually splits from
the bottom for large enough $|A_3|$. The onsite potential $V$ is
now said to soften. By analogy with the biphonon
theory~\cite{AGRA}, the splitting branch (labelled by $\{2\}$ in
Figs.~\ref{fig5}(b-c)) is identified as being the energy of
biphonon states. Measuring the energy of the biphonon with
reference to the unbound two-phonon states for same momentum $q$,
a biphonon binding energy can be defined. The absolute value
minimum, with respect to $q$ of the biphonon binding energy is
called the biphonon energy gap. When the non-quadratic
contribution is not sufficient to open the biphonon gap, the gap
is found to close at the center of the lattice Brillouin zone (BZ)
(Fig.\ref{fig5}(c)). On the other hand, at the edge of BZ the
binding energy may remain comparable to the width of the single
phonon branch. Thus, a pseudogap occurs when the non-quadratic
energy has the same magnitude as inter-site coupling. Similar
pseudogaps have been found in other quantum
lattices~\cite{Papanicolaou,Dorignac}. In the pseudogap regime,
biphonon excitations exist only at the edge of BZ while they
dissociate into unbound phonon pairs at center of the zone. In
Figs.~\ref{fig5}(b-c), it is noteworthy that the hybridization
between the biphonon states and the two-phonon states modifies
substantially the biphonon branch curvature and consequently the
quantum mobility of the biphonon. Such a feature can not be
analyzed accurately within the effective Hubbard model for bosons
since it neglects some energy terms that do not conserve the boson
number. The same remark holds for the single phonon dispersion law
which takes the form $(1+C cos(q a_0))$ (with arbitrary energy
units) in the effective boson model instead of the well-known
$\sqrt{1+2 C cos(q a_0)}$, calculated in the harmonic
approximation for similar parameters. The two formulas diverge
with increasing $C$, i.e. when the quantum hybridization is
amplified. This disagreement is due to the boson product operators
such as $(a^+_{i+1} a^+_{i})$ that are neglected in the effective
boson model, in contrast to the KG model and its harmonic
approximation.

\begin{figure}
\onefigure[width= 8cm]{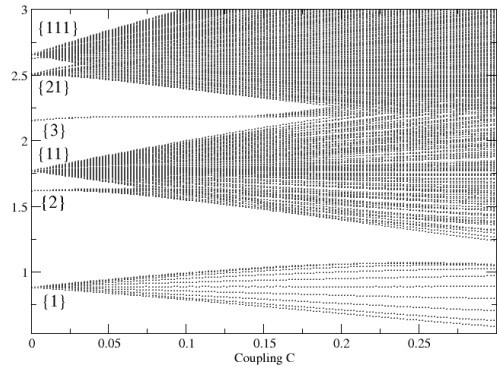}
\caption{Plot of
the energy spectrum versus dimensionless coupling $C$, for a 1D
quantum chain composed of $S=19$ cells with model parameters
$A_3=0.13$ and $A_4=0.01$. The energy unit is same as in
Fig.~\ref{fig5}. Labels are explained in the text.} \label{fig7}
\end{figure}

Other computations have been performed with different model
parameters to complete our results. The curvature of the biphonon
branch depends on the sign of $C$ provided the biphonon gap is
large. Otherwise, the hybridization between bound and unbound
phonons tends to impose the same curvature as the two-phonon
band(Fig.\ref{fig5}(c)). The biphonon band curvature thus results
from the interplay between the biphonon tunneling and the
hybridization with unbound phonons. In Fig.\ref{fig5}(b), the
triphonon energy branch (labelled by $\{3\}$) which splits from
the three-phonon band (not plotted) can also be noted with a
similar behavior as the biphonon branch.
For the same nonlinear coefficients as in Figs.\ref{fig5}(a-b),
varying artificially $C$ to the trivial case $C=0$ (see
Fig.~\ref{fig7}) demonstrates that the isolated branches, $\{1\}$,
$\{2\}$ and $\{3\}$ coincide with the energies of Bloch waves
$B_{[\alpha,0,...]}$ with a single onsite excitation $\alpha =1,2$
and $3$, respectively. The branch tag  corresponds to the
excitation level $\{\alpha\}$ of the coinciding Bloch waves at
$C=0$. The reason for the isolated branches is the anharmonicity
of $h_i$ eigenvalues $\gamma(\alpha>1)$ which implies some energy
gaps at $C=0$ that persist when $C\neq 0$. In Fig.~\ref{fig7}, we
note the bands associated to unbound phonon pairs, to unbound
three-phonon states and to unbound states composed of a biphonon
and a single phonon. These states are labelled by multiple tags
$\{\alpha, \beta, ...\}$ that correspond to Bloch waves
$B_{[\alpha,...,\beta...]}$ at $C=0$.

Since our calculations are performed with large enough lattices to
avoid boundary effects, there is a perfect agreement between the
energy spectrum computed with $S=33$ (Fig.~\ref{fig5}(a-b)) and
with $S=19$ (Fig.~\ref{fig7}) for the same parameters. In
Fig.~\ref{fig7}, the band widths of phonon bound states increase
with $C$ much slower than for unbound phonons. The branches of
biphonon $\{\alpha=2\}$ and triphonon $\{\alpha=3\}$ are found to
merge with the unbound phonon bands above a certain threshold
$C_\alpha$. For $C<C_\alpha$, the $\alpha^{th}$ branch is
separated from the rest of the spectrum by gaps that open over the
whole BZ whereas around $C\approx C_\alpha$, only a pseudogap
separates partially that branch from a certain unbound phonon
band. The $C_\alpha$ threshold depends on both $A_3$ and $A_4$ and
it is different for each phonon bound state. When $C$ is much
larger than $C_2$, the biphonon gap is closed then only gaps of
high order phonon bound states (such as triphonon) may be opened
in the excitation spectrum.
The results reported in Fig.~\ref{fig7} are qualitatively similar
to the Raman analysis~\cite{Mao} of the high pressure molecular
solid $H_2$ which shows a pressure-induced bound-unbound
transition of the so called bivibron, around $25$ GPa (compare
Fig.$10$ in Ref.~\cite{Mao} and the two-phonon energy region in
Fig.~\ref{fig7}). In the framework of our model, the pressure
variation of experiments~\cite{Mao} can be simulated by a change
of coupling integral $C$ due to the fact that neighboring
molecules are moved closer together because of the external
pressure. The increase of $C$ induces a bound-unbound transition
of the biphonon at $C=C_2$. The pressure variation of $C$ as well
as of other KG model parameters could inform us on how atomic
potentials depend on the inter-atomic distances (that are known
from diffraction measures). A biphonon is also exhibited by the
internal stretching~\cite{Jakob} of $CO$ molecules adsorbed on the
surface $Ru(111)$. There, the coverage of the surface can also be
thought to change the coupling integral $C$, and a bound-unbound
transition could thus be expected at high surface coverage. Yet,
that assumption is seemingly inapplicable according to a recent
theoretical work~\cite{Pouthier} which invoked a {\it coverage
dependant anharmonicity} which would originate in a chemical
modification of the intramolecular $CO$ potential due to
surrounding molecules.

\begin{figure}
\onefigure[width= 13cm]{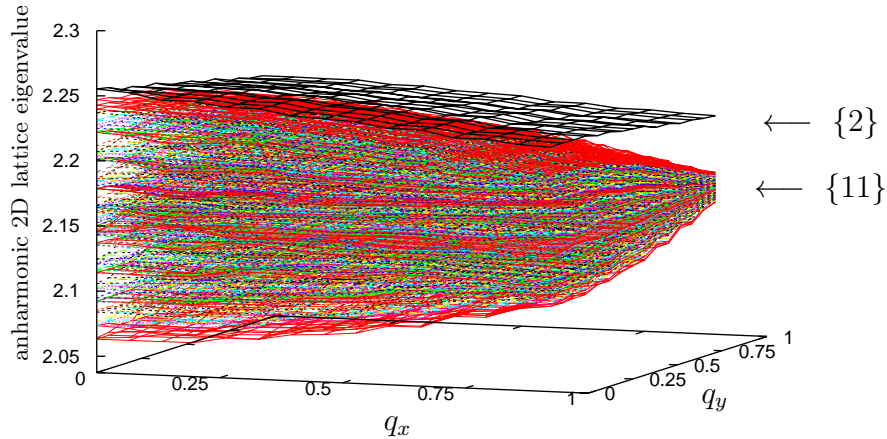} \caption{
Two-phonon spectrum for a 2D square lattice composed of
$S=13 \times 13$ cells and for the model parameters $C=0.025$,
$A_3=0$ and $A_4=0.025$. Energy unit and tags are same as in
Fig.~\ref{fig5}. The color picture (easier to read) is available online.}
\begin{picture}(100,1)(0,0)
\put(310,105){\makebox(0,0){\large $q_y $}}
\put(190,95){\makebox(0,0){\large $q_x$}}
\put(350,210){\makebox(0,0){\large $\longleftarrow \ \{2\} $}}
\put(350,185){\makebox(0,0){\large $\longleftarrow \ \{11\}$}}
\put(50,120){\begin{rotate}{90}
{\small anharmonic 2D lattice eigenvalue}
\end{rotate}}
\end{picture}
\label{fig9}
\end{figure}

Finally, for a 2D lattice, a biphonon pseudogap is found to open
around $q=1/2 [11]$ (see Fig.~\ref{fig9}). As we chose a set of
nonlinear parameters such that the local potential $V$ hardens,
the biphonon energy is higher than for two-phonon instead of lower
when $V$ softens (see Fig.~\ref{fig5}). Aside from
that point, the binding energy vanishes at center of BZ whereas
the width of pseudogap at the edge of BZ has same order as the
phonon branch width. That likeness between 1D and 2D spectra shows
that the pseudogap is a consistent property of lattices where
nonlinearity is comparable to the inter-site coupling, whatever
the dimensions are. For a 3D lattice, the biphonon pseudogap is
expected to open around $q=1/2 [111]$.\\

To summarize, the energy spectrum of the KG lattice has been
considered with numerics. The KG Hamiltonian provides a realistic
modelling of both quantum hybridization and pairing of phonon
states. Through our computing method, the spectral features of
phonon bound states can be related unambiguously to non-quadratic
terms of the potential energy. We predict that when those terms
are not strong enough to separate the spectral resonances of bound
and unbound phonon states, the pseudogaps of the biphonon,
triphonon or even higher order of phonon bound states, are
potential signatures of nonlinearity. The biphonon pseudogap has
been proved to open systematically at the edge of the first
Brillouin zone.

\acknowledgments

I gratefully acknowledge financial support from Trinity College
and an EC network grant on ``Statistical Physics and Dynamics of
Extended Systems''. Many thanks are addressed to Robert S. MacKay
and Serge Aubry.


\end{document}